\documentclass[12pt]{article}

\usepackage{amssymb}
\def\labelmark{}
\def\void{}

{\ifx\void\labelname\def\junk{\end{displaymath}}
\else\def\junk{\end{eqnarray}}\fi\junk\labelmark\def\labelname{}}

\newcommand{\bra}{\begin{array}}
\newcommand{\era}{\end{array}}
\newcommand{\beq}{\begin{equation}}
\newcommand{\eeq}{\end{equation}}
\newcommand{\bqn}{\begin{eqnarray}}
\newcommand{\eqn}{\end{eqnarray}}

\font\mybb=msbm10 at 12pt \def\bb#1{\hbox{\mybb#1}}   \font\mybbi=msbm10 at 9pt
\def\bbi#1{\hbox{\mybbi#1}}

\def\BC{\bb C}
\def\_\BC{\bbi C}

\newcommand{\om}{\omega}

\newcommand{\be}{\beta}

\newcommand{\pa}{\partial}
\newcommand{\al}{\alpha}

\newcommand{\del}{\delta}
\newcommand{\lga}{\longrightarrow}

\newcommand{\da}{\dagger}

\newcommand{\ov}{\over}

\newcommand{\lb}{\label}

\newcommand{\NP}[1]{ {\it Nucl.~Phys.} {\bf #1}}

\newcommand{\PRL}[1]{ {\it Phys.~Rev.~Lett.} {\bf #1}}

\newcommand{\JMP}[1]{ {\it J. Math.~Phys.} {\bf #1}}

\begin{document}
\begin{titlepage}
\setcounter{page}{1}
\renewcommand{\thefootnote}{\fnsymbol{footnote}}

\begin{flushright}
UCDTPG 05-01\\
hep-th/0505095
\end{flushright}

\vspace{6mm}
\begin{center}

{\Large\bf Quantum Hall Effect on Higher Dimensional Spaces}

\vspace{6mm}

{\small\bf Ahmed Jellal\footnote{ajellal@ictp.trieste.it}\footnote{Junior associate to ICTP}}

\vspace{4mm}

{\small \em Theoretical Physics Group,  
Faculty of Sciences, Chouaib Doukkali University},\\
{\small\em Ibn Ma\^achou Road, P.O. Box 20, 24000 El Jadida,
Morocco}\\

{\em and} \\

{\small \em Abdus Salam International Centre for Theoretical Physics},\\
{\small\em  Strada Costiera 11, 34014 Trieste, Italy}\\

\vspace{5mm}
\begin{abstract}

We analysis the quantum Hall effect exhibited by a
system of particles moving in a higher dimensional space. This
can be done by considering particles on the Bergman ball
$\sf{\bb{B}_{\rho}^d}$ of radius $\rho$ in the presence of an external
magnetic field $B$ and investigate its basic features. Solving
the corresponding Hamiltonian to get the energy levels as well as
the eigenfunctions. This can be used to study quantum Hall effect of
confined particles in the lowest Landau level
where density of particles and two point functions are calculated. We take advantage 
of the symmetry group of the Hamiltonian on $\sf{\bb{B}_{\rho}^d}$ to make link to the
Landau problem analysis on the complex projective spaces ${\sf CP^d}$. 
In the limit $\rho\lga\infty$, our analysis
coincides with that corresponding to particles on the flat geometry $\sf{\bb{C}^d}$.
This task has been done for $d=1, 2$ and finally
for the generic case, i.e.  $d \geq 3$.

\end{abstract}
\end{center}
\end{titlepage}

\section{Introduction}

Quantum Hall effect (QHE)~\cite{prange} is a fascinating subject
because not only of the precise quantized Hall conductivity but
also of its relationships to 
other subjects of theoretical physics and mathematics.
Since it appearance, the QH committee was and is still
considering particles constrained to move in two-dimensional
space. This is true because of the experiment evidences and
obligations.

More recently, people are talking about higher dimensional
QHE on different manifolds. One may ask how QHE can be realized 
on these spaces? The answer to
this question was done in 2001 from a theoretical point of view
where an embedding of QHE on 4-dimensional space was achieved
by Hu and Zhang~\cite{zhang}. The key point was to generalize
the Hall current from $SO(3)$ two-sphere $\sf{S^2}$  to
four-sphere $\sf{S^4}$ of the invariant group $SO(5)$.
 This is not the end, 
but it appeared very exciting and interesting works on the
subject. Indeed, QH droplets were considered on complex projective
spaces $\sf{CP^d}$~\cite{karabali1} where the wavefunctions were obtained and the
incompressibility of the Hall liquid was shown. 
Also based on  $\sf{CP^d}$, a
 relation between QHE and
the fuzzy spaces was discussed~\cite{knr}. 
Other important developments on the subject can be
found in~\cite{group0}. The common feature in these works is to
generalize the Landau problem on different higher dimensional manifolds. This is
because the Landau problem is the cornerstone of QHE.

On the other hand, investigating the Laplacian operator properties
on different manifolds have attracted several authors.
Many interesting results on this subject have been
reported by mathematicians. For instance, by considering
Laplacian on the ball $\sf{\bb{B}^d}$ with unit radius in different
dimensions the spectral theories were investigated.
For early work with $d=1$, for example one may
see~\cite{elstrodt,patterson}. In the case where $d\geq 2$,  we
refer to old work~\cite{folland} and the recent papers
can be found in~\cite{group1}.
Very recently, by taking into account
of radius as an additional degree of freedom
Ghanmi and Intissar~\cite{gintissar}
analyzed the same problem 
on the   
ball $\sf{\bb{B}_{\rho}^d}$ with $\rho$ positive real value.
Based on these works, we develop
our basic idea.

More specifically, we consider a system of
particles on the Bergman Ball $\sf{\bb{B}_{\rho}^d} $
and investigate its basic features.
This can by done by generalizing the Landau problem on the
plane to higher dimensional space $\sf{\bb{B}_{\rho}^d}$ and getting
its spectrum. The generalized Hamiltonian is invariant under
the group $SU(d,1)$ as well as $U(d)$. This will be used
to discuss a link to the Karabali and Nair work~\cite{karabali1}
on $\sf{CP^d}$. 
We  analytically start by analyzing the problem on
the disc that is $\sf{\bb{B}_{\rho}^1}$. This is isomorph to the
Poincar\'e half plane $\sf{\bb{H}}$ and therefore one may use the
Cayley transform to get the  $\sf{\bb{B}_{\rho}^1}$ spectrum from that
of particles living on $\bb{H}$. As we will see, the Landau problem on $\sf{\bb{B}_{\rho}^1}$
can be solved straightforwardly. 
This allows us to study particles
confined in the lowest Landau level (LLL) and discuss QHE
in terms of  the density of particles and the incompressibility
condition. This will be done after building the wavefunction
for the fully occupied state $\nu=1$.
Algebraically we make contact with particles moving in two-sphere $\sf{S^2}$,
which is $\sf{CP^1}$. This will be the first overlapping to~\cite{karabali1}. 
Subsequently, we treat 
the same problem but on  the ball $\sf{\bb{B}_{\rho}^2}$ by doing the same
job as for the disc and looking for link to $\sf{CP^2}$.
Finally, we consider the generic case
and this will be done by generalizing $\sf{\bb{B}_{\rho}^1}$ to higher dimensions. 
Also QHE of particles
in LLL will be investigated and link to $\sf{CP^d}$ will be emphasized. 
In all case we will refer to the flat geometry limit, which corresponds to
$\rho$ goes to infinity.

In section 2, for the necessity we review the Landau problem
on the plane and give its connection to QHE of particles in LLL. In section 3,
 we consider particles living on $\bb{H}$ in magnetic field $B$
and present the group theory approach to get the spectrum. This
can be done by realizing the appropriate group and exploiting its
representation. This will be used in section $4$ to discuss the link to $\sf{CP^1}$
after solving the Landau problem on the disc and discussing QHE in
LLL. This task will be done by analyzing
the Landau problem
on  $\sf{\bb{B}_{\rho}^2}$ and compared to $\sf{CP^2}$ in section 5.
We generalize our results to higher
dimensional space $\sf{\bb{B}_{\rho}^d}$ in section 6. 
We conclude and give some
perspectives in the final section.

\section{Flat geometry } 

Particles of mass $m$ living on the plane in an uniform
magnetic field $B$ is very interesting problem. Since it has much
to do with many areas of theoretical physics and in particular the
QHE subject. This is because of its
basic properties and the most important is its projection to 
LLL where particles try to be near the
minimal potential. This can be caused by applying a strong $B$
that generates a gap, which pushes particles to be in LLL. This
gives a natural example of the non-commutative geometry and
therefore a connection to beautiful theories like Laughlin
approach~\cite{laughlin}.

\subsection{Particle in the plane} 

For the necessity, we review the results of the Landau problem
on the plane. One particle Hamiltonian can be written as
\begin{equation}\lb{1hp}
H^{\sf F}= {1\over 4}\left\{-4\pa_z\pa_{\bar{z}} + B
\left(z\pa_{z}-\bar{z}\pa_{\bar{z}}\right)- \left({B\over
2}\right)^2|z|^2\right\}
\end{equation}
in the complex plane $(z,\bar{z})$ and the symmetric gauge
\begin{equation}\lb{gp}
A^{\sf F}={B\over 2}\left(y,-x\right).
\end{equation}
${\sf F}$ refers to Landau problem on the flat surface.
The unit system $(c,e,\hbar)$ is chosen to
be equal to $1$ and $m=2$. 
Without loss of generality, through this paper we assume that $B$ is strictly positive.

The spectrum can be obtained by
diagonalizing $H^{\sf F}$ in terms of creation and annihilation operators. 
These are
\beq
\lb{ob}
a^\da =2\pa_{\bar z}+{B\ov 2}{z}, \qquad
a =-2\pa_{z}+{B\ov 2}{{\bar z}}
\eeq
which satisfy the commutation relation
\beq\lb{cr}
\left[a^\da, a\right] = 2B.
\eeq
These can be used to write $H^{\sf P}$ as
\beq\lb{hca}
H^{\sf F} = {1\over 8} \left( a^\da a + a a^\da  \right).
\eeq
It is easily to see that the energy levels are  
\begin{equation}\lb{1elp}
E_l^{\sf F}= {B\over 4}\left(2l+{1}\right),\qquad l=0,1,2 \cdots
\end{equation}
and the eigenstates as well
\begin{equation}\lb{1ewfp}
|l\rangle^{\sf F}= \frac{1}{\sqrt{(2B)^l l!}}\left(a^{\da}\right)^l|0\rangle
\end{equation}
where $|0\rangle$ is the vacuum such as $a|0\rangle =0$.

The previous analysis can be generalized to 
many-particles system described by the total Hamiltonian 
\begin{equation}\lb{nhp}
H^{\sf F}_{\sf tot}={1\over 4} \sum_{i=1}^N\left\{-4\pa_{z_i}\pa_{\bar{z}_i}
+ B \left(z_i\pa_{z_i}-\bar{z}_i\pa_{\bar{z}_i}\right)-
\left({B\over 2}\right)^2|z_i|^2\right\}
\end{equation}
where the total energy is $N$ copies of (\ref{1elp}) and the eigenvalues
is basically the tensorial product of $N$ those given in (\ref {1ewfp}).

\subsection{LLL and QHE} 

The connection to QHE can be investigated by considering many
particles confined in LLL. 
We start by analyzing the spectrum of one particle, which
corresponds to $l=0$ in the above study. In this case, 
the complex space becomes non-commutative and  
we have
\beq\lb{nc}
\left[z, \bar{z} \right] = {2\over B}. 
\eeq
This is similar to the relation~(\ref{cr}). Therefore, by analogy to the 
previous analysis for one particle, we can  
define a Hamiltonian in LLL as
\beq\lb{lllh}
H^{\sf F}_{\sf LLL} = {1\over 8} \left( b^\da b + b b^\da  \right)
\eeq
where the  creation and annihilation operators
can be defined as
\beq
b^\da=z, \qquad  b=\bar{z}.
\eeq
It is clear that the spectrum can be read as
\begin{equation}\lb{elll}
\bra{ll}
E_{n,\sf{LLL}}^{\sf F}= {B\over 4}\left(2n+{1}\right),\\
|n\rangle_{\sf{LLL}}^{\sf F}= \sqrt{{B^n\over 2^n n!}}\left(a^{\da}\right)^n|0\rangle.
\era
\end{equation}
It is convenient to project the states on 
the complex plane $(z,\bar{z})$. Doing this to get the wavefunctions
\beq
\psi(z,\bar{z})={\sf{const}}\, z^n \exp\left(-{B\over 4}|z|^2\right)
\eeq

Let us consider $N$-particles in LLL, which of course means that all $l_i=0$
with $i=1,\cdots,N$ and each $l_i$ corresponds to the spectrum~(\ref{1elp}--\ref{1ewfp}).  
The total wavefunction
can be written in terms of the Slater determinant. This is
\begin{equation}\lb{nwps}
\psi(z,\bar{z})= \epsilon^{i_1 \cdots i_N} z_{i_1}^{n_1} \cdots z_{i_N}^{n_N}
\exp\left(-{B\over 4}\sum_i|z_i|^2\right)
\end{equation}
where $\epsilon^{i_1 \cdots i_N}$ is the fully
antisymmetric tensor and ${n_i}$ are integers. It is relevant to write this
wavefunction as Vandemonde determinant. We have  
\begin{equation}\lb{nwp}
\psi(z,\bar{z})= {\sf{const}}\,  \prod_{i,j}\left(z_i-z_j\right)
\exp\left(-{B\over 4}\sum_i|z_i|^2\right).
\end{equation}
This can be interpreted by remembering the Laughlin wavefunction 
\begin{equation}\lb{lw}
\psi_{\sf Laugh}^m(z,\bar{z})=  \prod_{i,j}
\left(z_i-z_j\right)^m\exp\left(-{B\over 4}\sum_i|z_i|^2\right).
\end{equation}
It is well known that it has many interesting features and good ansatz to describe
the fractional QHE at   the filling factor $\nu ={1\over m}$. It is clear that~(\ref{nwp}) 
is nothing but the first Laughlin state that corresponds to $\nu=1$. Actually,~(\ref{nwp})  
is describing the first quantized Hall plateau of the integer QHE. 
Note that~(\ref{lw}) can also be written as 
\begin{equation}\lb{lw2}
|m\rangle= \left\{\epsilon^{i_1 \cdots i_N} z_{i_1}^{n_1} \cdots z_{i_N}^{n_N}\right\}^m
|0\rangle.
\end{equation}

Before closing this section, we say some words about the filling factor because of
its relevance in 
the QHE world. In the unit system, this is
\begin{equation}\lb{ffp}
\nu= {2\pi {\cal{N}}\over B}
\end{equation}
where  ${\cal{N}}$ is the density of particles
\beq\lb{pdp}
{\cal{N}}={N\over S}
\eeq
and $S$ is the plane surface.
It is obvious that to obtain $\nu=1$, ${\cal N}$ should be equal to the finite quantity 
${B\over 2\pi}$. The QHE world tells us that $\nu$~(\ref{ffp}) must be quantized and reads as 
\begin{equation}\lb{qffp}
\nu= {N\over N_{\phi}}.
\end{equation}
This can be either fractional or integer, it depends to what the QHE kind
is involved. $ N_{\phi}$ is the number of the quantum flux 
\begin{equation}\lb{fp}
\Phi^{\sf F}= \int_{\sf{plane}} Bdxdy
\end{equation}
per unit of the flux $\Phi_0^{\sf F}={he\over c}$, which equal to $1$ in our 
choice of unit.
From this, one can learn that in getting the
QHE the magnetic field should be 
quantized. Note that $ N_{\phi}$ plays a crucial role since it
determines also the degree of degeneracy of Landau levels.

\section{Poincar\'e half plane $\sf{\bb{H}}$}

It will be clear in the next that the Poincar\'e half plane $\sf{\bb{H}}$
is isomorph to the ball $\sf{\bb{B}_{\rho}^1}$ . For this reason, we start
by treating the quantum mechanics of one particle living on  $\bb{H}$
in the presence of a magnetic field. This isomorphism can be used to
analysis the Landau problem on  $\sf{\bb{B}_{\rho}^1}$. 
Our study will be done analytically but we focus on the group theory approach
since it will involved in the next.
$\bb{H}$ is defined by
\begin{equation}\lb{php}
{\sf{\bb{H}}}=\{z=x+iy\in \bb{C}, y>0\}
\end{equation}
 and endowed with the metric
\begin{equation}\lb{phpm}
ds^2= {\rho^2\over y^2}\left(dx^2 + dy^2\right).
\end{equation}
The corresponding measure is
\begin{equation}\lb{phpme}
d\mu= {\rho^2\over y^2} dx dy.
\end{equation}

\subsection{Hamiltonian formalism}

One particle Hamiltonian $H^{\sf P}$ on  the Poincar\'e half plane $\sf{\bb{H}}$ can be derived
from the Laplace-Beltrami operator $H^{\bf LB}$ of a
particle of mass $m$ on a Riemann surface of metric $g_{ab}$ with a monopole
field~\cite{iengo}. This operator is given by
\begin{equation}\lb{lbop}
H^{\bf LB}= {1\over 2m } {1\over \sqrt{g}} p_a\left(\sqrt{g} g^{ab}\right) p^b
\end{equation}
where $p_a$ is a covariant derivative and $g$ is the metric determinant.
To obtain $H^{\sf P}$, first one can fix the gauge
\begin{equation}\lb{gauge}
A^{\sf P}=B\rho^2\left(-{1\over y}, 0\right)
\end{equation}
and second take into account of the associate metric~(\ref{phpm}). 
Doing these and setting $m=2$,~(\ref{lbop}) becomes
\begin{equation}\lb{phph}
H^{\sf P}={y^2\over\rho^2} \pa_z\pa_{\bar{z}}+
{i\over 2}By\left(\pa_z+\pa_{\bar{z}}\right) +{B^2\rho^2\over 4}.
\end{equation}
 ${\sf P}$ reflects the Landau problem on  $\sf{\bb{H}}$.~(\ref{phph})
generalizes the Hamiltonian~(\ref{1hp}) and 
can be recovered by taking the limit $\rho\lga\infty$.

There are tow possibilities to obtain the  spectrum of $H^{\sf P}$. This can
be done either algebraically or analytically. As it will be clear soon, the first is based on
the unitary irreducible representation of  $SL(2,\bb{R})$ that is the invariant
group of $H^{\sf P}$  but the second
is related to the eigenvalue equation. This is
\begin{equation}\lb{pee}
H^{\sf P} \Psi^{\sf P} = E^{\sf P} \Psi^{\sf P}.
\end{equation}
It can be solved on the compact manifold to get the eigenvalues
\begin{equation}\lb{pev}
E^{\sf P}_l = {1\over \rho^2} \left[\left(l+{1\over 2} -b\right)^2
+ b^2 +{1\over 4} \right]
\end{equation}
where $b={\rho}^2 B$ and the condition $0\leq l < b-{1\over 2}$ must be fulfilled.
The corresponding eigenfunctions are given by
\begin{equation}\lb{pes}
\Psi^{\sf P}_{l,k}(x,y)=
\sqrt{l! \left(2b-2l-1\right)\over 4\pi\rho^2 |k|\Gamma\left(2b-l\right)}
\exp\left(-ikx-|k|y\right)
\left(2|k|y\right)^{b-l} L_l^{2b-2l-1} \left(2|k|y\right)
\end{equation}
where
$L_{\al}^{\be}(x)$ is the generalized Laguerre functions
and $k$ is such as
\begin{equation}\lb{pact}
L_2 \Psi^{\sf P}_{l,k}(x,y) = -k \Psi^{\sf P}_{l,k}(x,y).
\end{equation}
$L_2=-\pa_x$ is a generator of the group $SL(2,\bb{R})$, see the next 
subsection.   
This shows that~(\ref{pes}) are degenerate wavefunctions. 
Note in passing that the present system has also a continue part, 
more detail can be found
in~\cite{comtet}. In the flat limit, the above spectrum coincides with that
obtained for one particle in section 2.

\subsection{Invariant group}

The system described by the Hamiltonian~(\ref{phph}) on the space $\sf{\bb{H}}$ 
can be analyzed by making use of
the group theoretical technology. This can be done by noting that $H^{\sf P}$ is invariant
under the group $SL(2,\bb{R})$. 
Its generators can be mapped in terms of 
the phase space variables of $H^{\sf P}$ such as
\beq\lb{gene} \bra {lll}
L_1=-i\left(z\pa_z+\bar{z} \pa_{\bar{z}}\right), \\
L_2=-\left(\pa_z+ \pa_{\bar{z}}\right), \\
L_3=i \left(z^2+\bar{z}^2\right)\pa_{\bar{z}} -ib\left(z-\bar{z}\right).
\era\eeq
They satisfy the $SL(2,\bb{R})$ commutation relations 
\begin{equation}\lb{sl2cr}
[L_1, L_2]=-iL_2,\qquad [L_1, L_3]=-iL_3, \qquad  [L_2, L_3]=2iL_1.
\end{equation}

Another group can be realized that is isomorph to  $SL(2,\bb{R})$.
This can be achieved by defining the generators
\begin{equation}\lb{jgene}
J_0={1\over 2}\left(L_2-L_3\right), \qquad J_1={1\over 2}\left(L_2+L_3\right), \qquad
J_2=L_1.
\end{equation}
One can check that they generate the unitary group $SU(1,1)$
\begin{equation}\lb{su11cr}
[J_0, J_1]= iJ_2,\qquad [J_0, J_2]=-iJ_1, \qquad  [J_1, J_2]= -iJ_0.
\end{equation}
An explicit relation between these generators and $H^{\sf P}$
can be fixed by introducing 
the Casimir operator of $SU(1,1)$. This is 
\begin{equation}\lb{casi}
C=J_0^2-J_1^2-J_2^2.
\end{equation}
Using~(\ref{gene}) and~(\ref{jgene}) to find
\begin{equation}\lb{cham}
H^{\sf P} = -{1\over 4 \rho^2} \left(C- b^2\right).
\end{equation}

From the last equation, one can learn that the spectrum of $H^{\sf P}$
can be obtained by handling the representation theory that 
corresponds to the operator $C$. 
To clarify this point, we
consider an unitary irreducible representation
of $SL(2,\bb{R})$ as eigenstates of $C$ as well as the compact
generator $J_0$~\cite{bargmann}. Otherwise, let us choose a basis 
$\{|j,m\rangle\}$ such as
\begin{equation}\lb{cact}
C|j,m\rangle = j(j+1) |j,m\rangle
\end{equation}
where $m$ is an eigenvalue of  $J_0$ 
\begin{equation}\lb{j0ev}
J_0 |j,m\rangle = m  |j,m\rangle.
\end{equation}
These results can be connected to those obtained by applying the 
analytical approach. This can be done by setting
\begin{equation}\lb{map}
j=l-b,\qquad
m=-l-b
\end{equation}
to end up with 
the derived energy levels as well as the eigenfunctions of $H^{\sf P}$.
More discussion about this issue can be found in~\cite{comtet}.

We have some remarks in order.
The generalization to $N$-particles without interaction
is immediate.
The quantum mechanics of particles on the plane
can be found by taking the limit $\rho\lga\infty$~\cite{comtet}.
The QHE exhibited by the present system on $\bb{H}$ was extremely studied in
reference~\cite{iengo}.
In conclusion let us emphasis that the above realization of $SU(1,1)$ 
will be helpful in the forthcoming analysis. With this, we discuss 
the connection to $\sf{CP^1}$, i.e. two-sphere. Also it can be generalized to 
the generic case, i.e. $d\geq 2$, and used to go deeply in investigating
other links to different spaces.

\section{Hyperbolic disc $\sf{\bb{B}_{\rho}^1}$}

Before considering particle living on higher dimensional
manifold in an external magnetic field and discuss the possibility to have QHE, we start
from the ball  $\sf{\bb{B}_{\rho}^1}$ of radius $\rho$ that is a particular case
and  corresponds to the disc. This is interesting task because
$\sf{\bb{B}_{\rho}^1}$ has much to do with
two-sphere $\sf{S^2}$ that is the complex projective space $\sf{CP^1}$.
This point will be clarified in the next study.
$\sf{\bb{B}_{\rho}^1}$ is 
\begin{equation}\lb{b1}
{\sf{\bb{B}_{\rho}^1}}=\left\{w\in \bb{C},|w|^2<\rho^2 \right\}
\end{equation}
and equipped with the Bergman-K\"ahler metric
\begin{equation} \lb{b1m}
\left(ds^{\sf 1}_{\rho}\right)^2={1\over \left(1-{|w|^2\over\rho^2}\right)^2}
dw\otimes d{\bar{w}}.
\end{equation}
In this space the integration can be done with respect to
the volume measure
\begin{equation} \lb{volb1}
d\mu_{\rho}^{\sf 1}(w)={1\over \left(1-{|w|^2\over\rho^2}\right)^2}
dm^{\sf 1}(w)
\end{equation}
where $dm^{\sf 1}(w)=dxdy$ is the Lebesgue form. Therefore
the inner product on  $\sf{\bb{B}_{\rho}^1}$ of two functions $\Psi_1$ and
 $\Psi_2$ in the Hilbert space reads as
 \begin{equation} \lb{ipb1}
\langle\Psi_1 |\Psi_2\rangle =\int_{\bb{B}_{\rho}^1}
d\mu_{\rho}^{\sf 1}(w) \bar{\Psi}_1 \times \Psi_2.
\end{equation}
In this section, we refer to $\sf{1}$ as $d=1$ and hereafter
$\rho\in ]0,\infty[$.

As we claimed before,
the Landau problem on the flat geometry
is the cornerstone of 
QHE. It may be a good task to treat the same problem
but at this time on the disc. This will allow us to investigate
the basic features of QHE and make contact with some relevant works. 
For instance, we propose to discuss a link to the results
obtained by considering particles on $\sf{CP^1}$.

\subsection{Spectrum}

To clarify the above ideas, we
consider the  Landau problem on the real hyperbolic disc
 and investigate its properties.
One particle Hamiltonian on $\sf{\bb{B}_{\rho}^1}$
can be obtained from the operator~(\ref{lbop}) associated to the metric~(\ref{b1m}).
This is
\begin{equation}\lb{dhf}
H^{\sf 1}=  \left(1- {|w|^2\over\rho^2} \right)\left\{-\left( 1 -
{|w|^2\over \rho^2}\right)
\pa_{w}\pa_{\bar{w}} -B\left(w\pa_{w}-\bar{w}\pa_{\bar{w}}\right)+
{B^2\over 4} |w|^2\right\}.
\end{equation}
The constant magnetic field is
 \begin{equation} \lb{magb1}
F^{\sf 1}=Bd\mu^{\sf 1}_{\rho}(w)=\left(\pa_{w}A_{w}^{\sf 1} -
\pa_{\bar{w}}A_{\bar{w}}^{\sf 1}\right)
dw\wedge d{\bar{w}}
\end{equation}
where $A_{w}^{\sf 1}$ and $A_{\bar{w}}^{\sf 1}$ are the complex
 compounds of the gauge $A^{\sf 1}$. These can be adjusted 
to get the corresponding magnetic flux as
\begin{equation} \lb{flb1}
\Phi^{\sf 1} =\int_{\sf{\bb{B}_{\rho}^1}} F^{\sf 1}=BS^{\sf 1}
\end{equation}
where  $S^{\sf 1}=\pi \rho^2$ is the area of $\sf{\bb{B}_{\rho}^1}$
mesured by the flat metric. Because of 
the Dirac quantization
$\Phi^{\sf 1}$ must be integral quantized. Therefore, we have 
\begin{equation} \lb{dqb1}
2k=B\rho^2
\end{equation}
where $k$ is integer value. This relation will be helpful in discussing QHE
generated from particles confined in LLL.

We make some comments about~(\ref{dhf}). This generalizes the Landau 
Hamiltonian~(\ref{1hp})
on the plane and can be recovered  
in the limit $\rho\lga\infty$. Since
$\sf{\bb{H}}$ is  isomorphic to  $\sf{\bb{B}_{\rho}^1}$, 
$H^{\sf 1}$ has something to do with $H^{\sf P}$. 
Effectively there is a relation between them and  
can be fixed by 
making use of the Cayley transformation to get
\begin{equation} \lb{ctr}
H^{\sf 1}= \left({z+i\over i-\bar{z}} \right)^{-k} H^{\sf P}
\left( {z+i\over i-\bar{z}} \right)^{k}
\end{equation}
such that $w$ and $z$ are governed by
\begin{equation} \lb{cmb1}
w = {z-i\over z +i}.
\end{equation}
Recalling that $w\in\sf{\bb{B}_{\rho}^1} $ and  $z\in\sf{\bb{H}}$. Equation~(\ref{ctr}) 
tells us that the eigenvalues and eigenstates of~(\ref{dhf})
can be derived from those corresponding to  $H^{\sf P}$. With this transformation 
the eigenvalue equation reads as
\begin{equation} \lb{eve}
H^{\sf 1}\Psi^{\rm 1}(w,\bar{w}) =\left( {z+i\over i-\bar{z}}  \right)^{-k} H^{\sf P}
\left[\left( {z+i\over i-\bar{z}}  \right)^{k} \Psi^{\rm 1}(w,\bar{w})\right].
\end{equation}

We present another way to get the analytical solution of the present problem
instead of using~(\ref{ctr}). This  first can be done   
by noting that $H^{\sf 1}$ is an elliptical operator on $\sf{\bb{B}_{\rho}^1}$,
the corresponding eigenfunctions
are $\bb{C}^{\infty}$--functions. They form a Hilbert space as
\beq
{\cal{H}} \left(\bb{B}_{\rho}^1\right) =\left\{ \Psi^{\sf 1}\in \bb{C}^{\infty}, \,
H^{\sf 1}\Psi^{\rm 1} (w,\bar{w}) = E^{\sf 1}\Psi^{\sf 1}(w,\bar{w})  \right\}.
\eeq
This property is important,  because now  
the eigenfunctions
$\Psi^{\rm 1} (w,\bar{w})$ can be expanded into  a spherical
series. This is
\begin{equation} \lb{cefb1}
\Psi^{\sf 1}(w,\bar{w}) = \sum_{p,q\geq 0}g_{p,q}(r^2) h^{p,q}(w,\bar{w}) 
\end{equation}
where the Fourier coefficient $g_{p,q}(r^2)$ are $\bb{C}^{\infty}$--functions
in $[0,\rho[$ with $|w|=r$. 
$h^{p,q} (w,\bar{w})$ are homogeneous harmonic polynomials on $\bb{C}^n$ and
of degree $p$ in $w$ and $q$ in $\bar{w}$~\cite{folland}. 
They form a space of restriction
to the boundary of the disk that is $U(1)$-irreducible representation.
Note that $h^{p,q} (w,\bar{w})$ are similar to those obtained by
Karabali and Nair~\cite{karabali1} in analyzing the algebraic structure of
particles on $\sf{CP^1}$.
These are
\begin{equation} \lb{hwf}
h^{p,q}(w,\bar{w}) = w^p \bar{w}^q. 
\end{equation}
On the other hand, one can easily check that~(\ref{cefb1})  are also
eigenfunctions of the angular momenta operator
\beq
L_w= w\pa_{w}-\bar{w}\pa_{\bar{w}} 
\eeq
with the eigenvalues
\beq
L_w \sum_{p,q\geq 0}g_{p,q}(r^2) h^{p,q}(w,\bar{w}) = 
 \sum_{p,q\geq 0} \left(p-q\right) g_{p,q}(r^2) h^{p,q}(w,\bar{w}).
\eeq

With the above equations and starting from 
\begin{equation} \lb{eeb1}
H^{\sf 1}\Psi^{\rm 1} (w,\bar{w}) = E^{\sf 1}\Psi^{\sf 1}(w,\bar{w})
\end{equation}
we end up with a differential equation for $ g_{p,q}\left(r^2\right)$. This is 
\begin{equation} \lb{eer}
\left[\left(1- {r^2\over\rho^2} \right)\left\{-\left( 1 -
{r^2\over \rho^2}\right) 
\left[r^2 \pa_{r^2}^2 + \left(1+p+q \right) \pa_{r^2}\right]
-B\left(p-q\right)+
{B^2\over 4} |w|^2\right\} -  E^{\sf 1} \right]g =   0.
\end{equation}
The solution of (\ref{eer}) can be worked out by setting 
\beq
g_{p,q}(r^2) = \left(1- u \right)^{{k\over 2}-l} F_{p,q}(u)
\eeq
and making the variable change ${r^2\over \rho^2}=u$. Injecting this in
(\ref{eer}), we find a hypergeometric
differential equation for $F(u)$ such as
\beq
\left\{ u(1-u) {d^2\over du^2 } + \left[p+q+1-\left(p+q+k-2l+1 \right)u \right]
-\left(q-l\right) \left(k+p-l\right) \right\} F_{p,q}(u) =0.
\eeq
It has a solution of the hepergeometric function type~\cite{magnus}
\beq
{}_2F_1\left( q-l, k+p-l; p+q+1; {r^2\over \rho^2} \right). 
\eeq
Therefore combining all, we obtain the eigenfunctions  
\begin{equation}
\Psi_l^{\sf 1}(w,\bar{w}) = \left(1-{r^2\over \rho^2}\right)^{{k\over 2}-l}
\sum_{p=0}^{\infty} \sum_{q=0}^{l}
{}_2F_1\left( q-l, k+p-l; p+q+1; {r^2\over \rho^2} \right)
h_{k,l}^{p,q} (w,\bar{w})
\end{equation}
associate to the eigenvalues 
\begin{equation}
E_l^{\sf 1}= {B\over 4}\left(2l+1\right) -{1\over \rho^2} l\left(l+1\right)
\end{equation}
where $l$ labels different Landau levels
of particle on the disc. It must satisfy the condition 
 $ 0\leq l< {k-1\over 2}$, recalling that $2k=B\rho^2$.

We introduce a physical quantity
since it has much to do with QHE. In particular, it can be used to
check the incompressibility condition of the present system in LLL.
This is the probability density,
which can be calculated to get
\begin{equation}\lb{nb1}
|\Psi_l^{\sf 1}|^2 = \sum_{p=0}^{\infty} \sum_{q=0}^{l}
Y_{k,l}^{1,p,q}
|h_{k,l}^{p,q}|^2
\end{equation}
where $Y_{k,l}^{1,p,q}$ is a function of the radius $\rho$ 
\begin{equation}\lb{yb1}
Y_{k,l}^{1,p,q} ={(l-q)!\rho^{2(1+p+q)} \over 2(k-1-2l)}
{\Gamma ^2(1+p+q) \Gamma(k-q-l) \over \Gamma(1+p+l){\Gamma}(k+p-l)}.
\end{equation}
We remark that only $|\Psi_l^{\sf 1}|^2$ and $|h_{k,l}^{p,q}|^2$
are complex coordinate dependent.

Since the above results generalize those of the Landau problem
on the flat surface, it is relevant to check the asymptotic behavior.
This can be achieved by sending
$\rho$ to infinity. Doing this to obtain the energy levels 
 \begin{equation}
E_l^{\sf F}= {B\over 4} \left(2l+1\right).
\end{equation}
This coincides exactly with (\ref{1elp}).
The corresponding wavefunctions are
\begin{equation}\lb{lwb1}
\Psi_l^{\sf F}(w,\bar{w}) = e^{-{B\over 4}r^2}
\sum_{p=0}^{\infty} \sum_{q=0}^{l}
{}_1F_1\left( q-l, p+q+1;  {B\over 2} r^2 \right)h_{l}^{p,q} (w,\bar{w}).
\end{equation}
They are analogue to the common eigenfunctions of the Hamiltonian~(\ref{1hp}) and the
angular momenta 
\beq
L_z=z\pa_{z}-\bar{z}\pa_{\bar{z}}.
\eeq
One may recall the formula~\cite{magnus}
\begin{equation}\lb{fgl}
{}_1F_1\left( -\mu, 1+\be;  x \right)=
{\Gamma(1+\be) \Gamma(1+\mu) \over \Gamma(\be+\mu+1)}
L^{\be}_{\mu}(x)
\end{equation}
where $L^{\be}_{\mu}(x) $ is the generalized Laguerre functions.

\subsection{QHE on $\sf{\bb{B}_{\rho}^1}$ }

Particles in  LLL are confined in a potential
that is grand enough to neglect the kinetic energy. This
produces a gap such that  particles are not allowed to jump to
the next level. LLL is rich and contains many interesting
features that are relevant in discussing QHE. 
With this, it is interesting to
consider our system on LLL. 

We keep particles in LLL and investigate their basic features. 
We start by giving the
 spectrum for one particle in LLL, which can be obtained 
just by fixing $l=0$ in the previous analysis. This gives 
the ground state
\begin{equation}\lb{gsb1}
\Psi^{\sf 1}_0(w) = \left(1-{r^2\over \rho^2}\right)^{k\over 2}
\sum_{p=0}^{\infty}
{}_1F_1\left( k+p, p+1; {r^2\over \rho^2} \right)
h_{k}^{p} (w)
\end{equation}
that has the energy
\begin{equation}\lb{0eb1}
E^{\sf 1}_0= {B\over 4}.
\end{equation}
This  value coincides with  that corresponds to  the
Landau problem on the plane. Actually 
$h_{k}^{p} (w)$ is a polynomial of degree $p$ in $w$.
The density in LLL reads as
\begin{equation}\lb{nlb1}
|\Psi_0^{\sf 1}|^2 = \sum_{p=0}^{\infty} 
{\rho^{2(1+p)}
\Gamma (1+p) \Gamma(k) \over 2(k-1) {\Gamma}(k+p)}
|h_{k}^{p}|^2.
\end{equation}

One particle spectrum in LLL can be generalized to
that for $N$-particles. It is obvious that the total energy is given by 
\begin{equation}\lb{N0eb1}
E_N^{\sf 1}= {NB\over 4}.
\end{equation}
The corresponding wavefunction can be constructed as the Slater determinant. This is 
\begin{equation}\lb{Ngsb1}
\Psi^{\sf 1}_N(w)= \epsilon^{i_1 \cdots i_N}
\Psi^{\sf 1}_{i_1}(w_{i_1})  \Psi^{\sf 1}_{i_2}(w_{i_2})
\cdots  \Psi^{\rm 1}_{N_1}(w_{i_N})
\end{equation}
where each  $\Psi^{\rm 1}_{i_j}(w_{i_j})$ has the form given
in~(\ref{gsb1}). This is similar to the wavefunction
(\ref{nwps}) on the plane and  corresponds to the filling 
factor $\nu=1$. Other similar Laughlin states can be obtained as 
we have in~(\ref{lw2}).

The definition (\ref{ffp}) tells us that the density of particle is an important
ingredient. To get QHE, this parameter should be kept constant
by varying  the magnetic field. For its relevance, we evaluate the
density by 
defining the number of particles as
\beq\lb{defN}
N =  \int_{\bb{B}_{\rho}^1} {\cal{N}}^{\sf 1}\left(w\right)
d\mu_{\rho}^{\sf 1}\left(w\right)= \pi \rho^2 {\cal{N}}^{\sf 1}_{0}.
\eeq
This gives
\begin{equation}
{\cal{N}}^{\sf 1}_{0}={N\over \pi \rho^2} = {BN\over 2 \pi k}.
\end{equation}
In the thermodynamic limit $N,\rho\lga \infty$, it goes to 
the finite quantity 
\beq
{\cal{N}}^{\sf 1}_{0} \sim {B\over 2\pi}. 
\eeq
This is exactly the density of particles on flat geometry and therefore
 corresponds to the fully occupied state 
 $\nu=1$.

In QHE the quantized plateaus come from the realization of an incompressible liquid.
 This property is important since it is related to the energy. It 
means that by applying an infinitisemal 
pressure to an incompressible system the volume remains unchanged~\cite{ezawa}. 
This condition can be checked for our system by
considering two-point function and integrating over
all particles except two.  This is 
\begin{equation}\lb{tpfb1}
I^{\sf 1}(w_{i_1},w_{i_2})= \int_{\bb{B}_{\rho}^1} 
d\mu_{\rho}^{\sf 1}\left(w_3,w_4,\cdots, w_N\right)
\left[\Psi^{\sf 1}_N(w)\right]^{*} \Psi^{\sf 1}_N(w).
\end{equation}
It is easy to see that $I^{\sf 1}(w_{i_1},w_{i_2})$ is
\begin{equation}\lb{2tpfb1}
I^{\sf 1}(w_{i_1},w_{i_2})\sim |\Psi_{0,i_1}^{\sf 1}|^2 |\Psi_{0,i_2}^{\sf 1}|^2-
|\left(\Psi_{0,i_1}^{\sf 1}\right)^*\Psi_{0,i_2}^{\sf 1} |^2.
\end{equation}
This can also be evaluated in the plane limit. Taking $\rho\lga\infty$, we obtain 
\beq
\lb{22tpfb1}
I^{\sf 1}(w_{i_1},w_{i_2})\sim e^{-{B\over 2}\left(r_1^2- r_2^2\right)}
\sum_{p_1,p_2\geq 0} \left(|w_1^{p_1}|^2  |w_2^{p_2}|^2 - |w_1^{p_1}w_2^{p_2} |^2 \right).
\eeq
This equation tells us that the probability of finding two particles at the same position is zero,
as should be.

\subsection{Two-sphere}

In $1983$, Haldane~\cite{haldane} proposed an approach to overcome
the symmetry  problem that brought by the Laughlin theory for the
fractional QHE describing by the wavefunction~(\ref{lw}) at the filling factor 
$\nu={1\over m}$. More precisely,~(\ref{lw}) is 
rotationally invariance due to the angular momenta but not 
transtionally. 
By considering particles living on two-sphere in a magnetic monopole,
Haldane formulated a theory that   possess all symmetries and 
generalizes the Laughlin proposal.
Very recently, Karabali and
Nair~\cite{karabali1} elaborated an algebraic analysis that
supports the Haldane statement and gives a more general results.
These developments on  $\sf{S^2}$ will be connect to our study on the
disc, i.e. $\sf{\bb{B}_{\rho}^1}$.

The link to two-sphere $\sf{S^2}$ can be done via the following considerations.
First,  $\sf{S^2}$ can be realized  on the disc as
\begin{equation}\lb{sp1}
\pa{\sf{\bb{B}_{\rho}^1}}={\sf{S^2}}=\left\{\om\in \bb{C}, |w| =\rho \right\}.
\end{equation}
It is basically the ball $\sf{\bb{B}_{\rho}^1}$ boundary. This tells that from the basic
features of $\sf{\bb{B}_{\rho}^1}$ one can derive those of $\sf{S^2}$. 
Second, $H^{\sf 1}$ is invariant on the
symmetric space 
\begin{equation}\lb{cp}
{SU(1,1)\over U(1)}.
\end{equation}
To get the complex projective space $\sf{CP^1}$ 
\beq
{\sf{CP^1}}\equiv {SU(2)\over U(1)}
\eeq
that is $\sf{S^2}$, one can use an analytic continuation
of $SU(1,1)$ to $SU(2)$ (as in the Weyl unitary trick for
groups).
It suggests that the obtained
spectrum on $\sf{\bb{B}_{\rho}^1}$ is similar to that 
of the Landau problem on the sphere except that our eigenfunctions
should be invariant under $ U(1)$.

Algebraically to make contact with 
our analysis on  $\sf{\bb{B}_{\rho}^1}$,
we refer to the symmetry group
(\ref{cp}) and report what Karabali and
Nair~\cite{karabali1} have done in terms
of our language. In this case, one 
can factorize the Hamiltonian 
$H^{\sf 1} $ as 
\begin{equation}\lb{fham}
H^{\sf 1} = -\left(D_+D_-+D_-D_+\right).
\end{equation}
At this stage, one may do recall to the generators of $SL(2,\bb{R})$. This can be done
by mapping the covariant derivatives $D_{\pm}$ as follows
\begin{equation}\lb{dop}
D_{+} = {1\over \rho} L_{2}, \qquad D_{-} = {1\over \rho} L_{3}
\end{equation}
and the relation must be fulfilled
\begin{equation}\lb{dcom}
\left[D_+, D_-\right] = -{B\over 2}.
\end{equation}
This and (\ref{dop}) fix the eigenvalue of the generator $L_1$ 
as $i{k\over 2}$. The Hamiltonian can be written in terms of
the Casimir and $J_2$ of the group $SU(1,1)$ as 
\begin{equation}\lb{lham}
H^{\sf 1} = -{1\over \rho^2}\left(C +J_2^2\right).
\end{equation}
It is clear that to get
the spectrum of   $H^{\sf 1}$,
one can use the representation theory where the eigenvalues of
$C$ are $j(j+1)$. This gives the energy levels for 
 \begin{equation}\lb{krel}
E_l^{\sf 1} = -{1\over \rho^2} \left[\left(l-{k\over 2}\right) \left(l-{k\over 2}+1\right)
-{k^2\over 4}\right]
\end{equation}
by fixing  
\beq
j=l-{k\over 2}
\eeq
and taking into account the Dirac quantization $2k=B\rho^2$. This leads to the same result
as we have derived analytically for one particle on the disc.
Therefore, one may use the Karabali and
Nair~\cite{karabali1} technology to make contact with our analysis on QHE.

In conclusion, the above study
suggests to go further and
get other link to the complex projective spaces in higher dimensions.
Next, we consider the ball   $\sf{\bb{B}_{\rho}^2}$ and
see what has common with $\sf{CP^2}$.

\section{Ball $\bb{B}_{\rho}^{\sf{2}}$ } 

We deal with another interesting case that is
the Landau problem on the ball $\sf{\bb{B}_{\rho}^2}$ and compare our results
to those derived by analyzing the same
problem on the  complex projective space  $\sf{CP^2}$~\cite{karabali1}. 
We start by defining $\sf{\bb{B}_{\rho}^2}$ as
\begin{equation}
\bb{B}_{\rho}^{\sf{2}}=\left\{w =(w_1,w_2)\in \bb{C}^2, |w|^2
= |w_1|^2 + |w_2|^2  <\rho \right\}.
\end{equation}
Its K\"ahler-Bergman metric is given by
\begin{equation}\lb{mb2}
\left(ds^{\sf 1}_{\rho}\right)^2={\rho^2\over\left(\rho^2-|w|^2\right)^2}
\sum_{i,j=1}^2\left[\left(\rho^2- w_i\bar{w}_j\right) \del_{ij} + w_i\bar{w}_j\right]
dw_i\otimes d{\bar{w}}_j.
\end{equation}
and the measure reads as
\begin{equation}\lb{vb2}
d\mu_{\rho}^{\sf 2}(w)={1\over\left(1-{|w|^2\over\rho^2} \right)^{3}}
dm^{\sf 2}(w)
\end{equation}
The volume of the ball $\bb{B}_{\rho}^{\sf{2}}$ measured by the
Euclidean
metric is
\beq 
S^{\sf  2}= {1\over 2}\pi^2\rho^4.
\eeq  
This is interesting because it will be used in calculating  the 
density of particles living on $\sf{\bb{B}_{\rho}^2}$. Of course the
indices ${\sf{2}}$ 
refers to $d=2$.

\subsection{Landau problem on $\sf{\bb{B}_{\rho}^2}$ }

We study the Landau problem on the ball $\sf{\bb{B}_{\rho}^2}$. This will be a generalization
to $d=2$ of our previous results obtained by treating the same problem on
the disc. Let us consider one particle living on $\sf{\bb{B}_{\rho}^2}$ in
an external magnetic field $B$. The Hamiltonian can be written as
\begin{equation}\lb{hb2}
H^{\sf 2}= \left(1- {|w|^2\over\rho^2}\right) 
\sum_{i=1}^2\left\{-\sum_{j=1}^2 \left( \del_{ij} +
{w_i\bar{w}_j\over \rho^2}\right)
\pa_{w_i}\pa_{\bar{w}_j} -B\left(w_i\pa_{w_i}-\bar{w}_i\pa_{\bar{w}_i}\right)+
{B^2\over 4} |w_i|^2\right\}.
\end{equation}
In this case, the limit $\rho\lga\infty$ corresponds to the Landau Hamiltonian
on the complex space of dimension $d=2$, i.e. $\sf{\bb{C}^{2}}$.

The spectrum can be obtained by solving the eigenvalue equation 
\begin{equation}
H^{\sf 2}\Psi^{\sf 2}=E^{\sf 2}\Psi^{\sf 2}.
\end{equation}
It can be worked out by applying the same method as we have done for the disc. 
This gives the energy levels 
\begin{equation}\lb{1elb2}
E_l^{\sf 2}= {B\over 4}\left(2l+2\right) -{1\over \rho^2} l\left(l+2\right)
\end{equation}
and the eigenfunctions 
\begin{equation}\lb{1ewfb2}
\Psi^{\sf 2}_{l}(w,\bar{w}) = \left(1-{|w|^2\over \rho^2}\right)^{{k\over 2}-l}
\sum_{p=0}^{\infty} \sum_{q=0}^{l}
{}_2F_1\left( q-l, k+p-l; p+q+2; {|w|^2\over \rho^2} \right)h_{k,l}^{p,q} (w,\bar{w})
\end{equation}
where
 $ 0\leq l< {k\over 2} -1$ and~(\ref{dqb1}) holds. 
Note that $h_{k,l}^{p,q} (w,\bar{w})$ form a space of restriction to 
$\pa\bb{B}_{\rho}^{\sf{2}}$ and have the same form as given in (\ref{hwf}) except that
$w$ has two-components. 
We remark that the spectrum is dimension $d=2$ dependent. This effect
will manifest in discussing QHE on $\sf{\bb{B}_{\rho}^2}$ and of course make 
difference with respect to the case $d=1$.

The probability density is  
\begin{equation}\lb{nb2}
|\Psi_l^{\sf 2}|^2 = \sum_{p=0}^{\infty} \sum_{q=0}^{l}
Y_{k,l}^{{\sf 2},p,q}
|h_{k,l}^{p,q}|^2
\end{equation}
where $Y_{k,l}^{{\sf 2},p,q}$ reads as
\begin{equation}\lb{yb2}
Y_{k,2}^{{\sf 2},p,q} ={(l-q)!\rho^{2(2+p+q)} \over 2(k-2l-2)}
{\Gamma ^2(2+p+q) \Gamma(k-q-l-1) \over \Gamma(2+p+l) {\Gamma}(k+p-l)}.
\end{equation}

The spectrum on the complex plane $\sf{\bb{C}^2}$ can be obtained 
in the limit $\rho\lga\infty$. 
This leads to the energy
 \begin{equation}
E_l^{\sf L}= {B\over 4} \left(2l+2\right)
\end{equation}
and the wavefunctions 
\begin{equation}\lb{lwb2}
\Psi_l^{\sf L}(w,\bar{w}) = e^{-{B\over 4}r^2}
\sum_{p=0}^{\infty} \sum_{q=0}^{l}
{}_1F_1\left( q-l, p+q+2;  {B\over 2} r^2 \right)h_{l}^{p,q} (w,\bar{w}).
\end{equation}

\subsection{LLL analysis }

By restricting to LLL we discuss  the QHE on $\sf{\bb{B}_{\rho}^2}$ 
in terms of the density
of particles and the incompressibility of our system. As usual
to be in LLL, we consider the ground state energy that is corresponds to 
$l=0$ in~(\ref{1elb2}). We have 
\begin{equation}\lb{0eb2}
E_0^{\sf 2}= {B\over 2}.
\end{equation}
This value is similar to that obtained by Karabali and Nair~\cite{karabali1}.
Its wavefunction can be obtained in the same way from~(\ref{1ewfb2})
\begin{equation}\lb{gsb2}
\Psi^{\sf 2}_0(w) = \left(1-{|\om|^2\over \rho^2}\right)^{k\over 2}
\sum_{p=0}^{\infty}
{}_1F_1\left( k+p, p+2; {|w|^2\over \rho^2} \right)
h_{k}^{p} (w).
\end{equation}
It has the density 
\begin{equation}\lb{nlb2}
|\Psi_0^{\sf 2}|^2 = \sum_{p=0}^{\infty} 
{\rho^{2(d+p)}
\Gamma (2+p) \Gamma(k-1) \over 4(k-1) {\Gamma}(k+p)}
|h_{k}^{p}|^2.
\end{equation} 

It is obvious that for $N$ particles in  LLL, the energy is
$N$ copies of $E_0^{\sf 2}$. This is
\begin{equation}
E_N^{\sf 2}= {NB\over 2}
\end{equation}
and its wavefunction can be built in similar way as we have done  
in~(\ref{Ngsb1})
\begin{equation}\lb{Ngsb2}
\Psi^{\sf 2}_N(w)= \epsilon^{i_1 \cdots i_N}
\Psi^{\sf 2}_{i_1}(w_{i_1})  \Psi^{\sf 2}_{i_2}(w_{i_2})
\cdots  \Psi^{\sf 2}_{N_1}(w_{i_N})
\end{equation}
where $\Psi^{\sf 2}_{i_j}(w_{i_j})$ is given
by~(\ref{gsb2}). 

The density of $N$-particles can be calculated by
using the area of $\sf{\bb{B}^{\sf 2}_{\rho}}$ and adopting a
definition of $N$ analogue to~(\ref{defN}). We have 
\begin{equation}
{\cal{N}}^{\sf 2}_0={2N\over \pi^2 \rho^4} = {NB^2\over 2\pi^2 k^2}.
\end{equation}
In the thermodynamics limit, we obtain
\beq
{\cal{N}}^{\sf 2}_0 \sim {1\over 2}\left({B\over \pi}\right)^2. 
\eeq
This agrees with 
Karabali and Nair~\cite{karabali1} result and conclusion. More precisely, 
it means that there
is no need to introduce an infinite internal degrees of freedom
as it has been done in~\cite{zhang}.

In similar fashion to the disc,
we can check the incompressibility of the system by evaluating
 two-point function. This is  
\begin{equation}\lb{tpfb2}
I^{\sf 2}(w_{i_1}^{\sf 2},w_{i_2}^{\sf 2})= \int_{\bb{B}_{\rho}^2} d\mu^{\sf 2}_{\rho}
\left(w_3,w_4,\cdots, w_N\right)
\left[\Psi^{\sf 2}_N(w)\right]^{*} \Psi^{\sf 2}_N(w).
\end{equation}
This can be integrated 
to obtain 
\begin{equation}\lb{2tpfb2}
I^{\sf 2}(w_{i_1}^{\sf 2},w_{i_2}^{\sf 2})\sim |\Psi_{0,i_1}^{\sf 2}|^2 |\Psi_{0,i_2}^{\sf 2}|^2-
|\left(\Psi_{0,i_1}^{\sf 2}\right)^*\Psi_{0,i_2}^{\sf 2} |^2.
\end{equation}
If we consider the limit $\rho\lga\infty$, we find
\beq
\lb{22tpfb2}
I^{\sf 2}(w_{i_1}^{\sf 2},w_{i_2}^{\sf 2})\sim 
e^{-{B\over 2}\left(r_1^2- r_2^2\right)}
\sum_{p_1,p_2\geq 0} \left(|w_1^{p_1}|^2  |w_2^{p_2}|^2 - |w_1^{p_1}w_2^{p_2} |^2 \right).
\eeq
It leads to the same conclusion as for the disc except that each ${w}_{i_j}$
has two components.

To close this section, we note that to make contact with 
the Landau problem
on  the complex projective space  $\sf{CP^{2}}$ one can use 
the theory group approach. This can be done
by considering the invariant symmetric space 
\beq
 {SU(2,1)\over U(2)}
\eeq
of the Hamiltonian $H^{\sf 2}$ and using an analytic continuation of
$SU(2,1)$ to $SU(3)$, in similar way as for the disc and 
$\sf{CP^{1}}$, to have a link to
\beq
{\sf{CP^{2}}} \equiv  {SU(3)\over U(2)}.
\eeq

\section{Spaces $\sf{\bb{B}_{\rho}^{d}}$}

We
generalize the task done in two last sections 
by considering
particles on  the Bergman ball $\sf{\bb{B}_{\rho}^{d}}$
of dimension ${d}$.
This is a generalization to higher
dimensional complex spaces, which can be achieved by
replacing the plane $\sf{\bb{C}}$ by the Hermitian complex space
 $\sf{\bb{C}^d} (d\geq 1)$ and the real hyperbolic disc
 by the Bergman complex ball of radius  $\rho >0$. We have
\begin{equation}
{\sf{\bb{B}_{\rho}^{d}}}=\left\{w =(w_1,w_2,\cdots,w_d)\in \bb{C}^d, |w|^2
= |w_1|^2 + |w_2|^2 + \cdots + |w_d|^2 <\rho \right\}.
\end{equation}
This is endowed with the K\"ahler-Bergman metric  
\begin{equation}\lb{mbn}
\left(ds^{\sf d}\right)_{\rho}^2={\rho^2\over\left(\rho^2-|w|^2\right)^2}
\sum_{i,j=1}^d\left[\left(\rho^2- w_i\bar{w}_j\right) \del_{ij} + w_i\bar{w}_j\right]
dw_i\otimes d{\bar{w}}_j.
\end{equation}
The measure is
\begin{equation}\lb{vbn}
d\mu_{\rho}^{\sf d}(w)={1\over\left(1-{|w|^2\over\rho^2} \right)^{d+1}}
dm^{\sf d}(w)
\end{equation}
where $dm^{\sf d}(w) = dw_1 dw_2 \cdots dw_d$ is the Lebesgue measure in
the flat geometry
 $\sf{\bb{C}^d}$. In general the surface of  $\sf{\bb{B}_{\rho}^{d}}$ 
measured by the flat metric is 
given by
\beq
S^{\sf d} = {\pi^{d\over 2}\rho^d\over \Gamma\left({d\over 2}+1\right)}
\eeq
where $\Gamma\left({d\over 2}+1\right)$ is the gamma function.

\subsection{Generalized Hamiltonian }

We start by considering the Hamiltonian $H^{\sf d}$ of one 
particle living on  $\sf{\bb{B}_{\rho}^{d}}$
in an external magnetic field $B$. This is the 
sum over ${d}$ of $H^{\sf 1}$ such as 
\begin{equation}\lb{hbd}
H^{\sf d}= \left(1- {|w|^2\over\rho^2}\right) 
\sum_{i+1}^d\left\{-\sum_{j=1}^d \left( \del_{ij} +
{w_i\bar{w}_j\over \rho^2}\right)
\pa_{w_i}\pa_{\bar{w}_j} -B\left(w_i\pa_{w_i}-\bar{w}_i\pa_{\bar{w}_i}\right)+
{B^2\over 4} |w_i|^2\right\}.
\end{equation}
In the limit $\rho\lga\infty$,  $H^{\sf d}$ goes to the Landau Hamiltonian
on the complex space $\sf{\bb{C}^d}$.

The spectrum of~(\ref{hbd}) can be obtained via two methods. The first one
based on the weighted Plancherel formula~\cite{gintissar}. The second is
related to the standard method that is the eigenvalue equation 
\begin{equation}
H^{\sf d}\Psi^{\sf d}=E^{\sf d}\Psi^{\sf d}
\end{equation}
This can solved in similar fashion as we have done for the disk. The obtained
results can be summarized as follows. With the condition  $ 0\leq l< {k-d\over 2}$
and~(\ref{dqb1}), 
the energy levels read as
\begin{equation}\lb{elbd}
E_l^{\sf d}= {B\over 4}\left(2l+d\right) -{1\over \rho^2} l\left(l+d\right)
\end{equation}
and the eigenfunctions are
\begin{equation}\lb{wfbd}
\Psi^{\sf d}_l(w,\bar{w}) = \left(1-{|w|^2\over \rho^2}\right)^{{k\over 2}-l}
\sum_{p=0}^{\infty} \sum_{q=0}^{l}
{}_2F_1\left( q-l, k+p-l; p+q+d; {|w|^2\over \rho^2} \right)h_{k,l}^{p,q} (w,\bar{w}).
\end{equation}
Recalling that 
$w$ has ${d}$-components. Again $h_{k,l}^{p,q} (w,\bar{w})$ form a space of restriction to
the boundary of $\bb{B}_{\rho}^{\sf d}$.
We note that both of energy and wavefunctions are dimensions $d$-dependent. 

The density of~(\ref{wfbd}) can be calculated to obtain
\begin{equation}\lb{nbd}
|\Psi_l^{\sf d}|^2 = \sum_{p=0}^{\infty} \sum_{q=0}^{l}
Y_{k,l}^{{\sf d},p,q}
|h_{k,l}^{p,q}|^2
\end{equation}
where $Y_{k,l}^{{\sf d},p,q}$ is 
\begin{equation}\lb{ybd}
Y_{k,l}^{{\sf d},p,q} ={(l-q)!\rho^{2(d+p+q)} \over 2(k-d-2l)}
{\Gamma ^2(d+p+q) \Gamma(k-d-q-l) \over \Gamma(d+p+l) {\Gamma}(k+p-l)}
\end{equation} 
It is clear that for  $d=1,2$ we recover the spectrum of the disc and  
$\bb{B}_{\rho}^2$, respectively.

We make contact with the flat geometry $\sf{\bb{C}^d}$ by sending 
 $\rho$ to infinity. We have the energy 
 \begin{equation}
E_l^{\sf L}= {B\over 4} \left(2l+d\right)
\end{equation}
and the corresponding wavefunctions 
\begin{equation}\lb{lwbd}
\Psi_l^{\sf L}(w,\bar{w}) = e^{-{B\over 4}r^2}
\sum_{p=0}^{\infty} \sum_{q=0}^{l}
{}_1F_1\left( q-l, p+q+d;  {B\over 2} r^2 \right)h_{l}^{p,q} (w,\bar{w}).
\end{equation}

\subsection{Particles in LLL on  $\sf{\bb{B}_{\rho}^d}$}

We generalize our former study on LLL to higher dimensions and
investigate the QHE basic features.
We start with one particle ground state  
\begin{equation}\lb{gsbd}
\Psi^{\rm B}_0(w) = \left(1-{|\om|^2\over \rho^2}\right)^{k\over 2}
\sum_{p=0}^{\infty}
{}_1F_1\left( k+p, p+d; {|w|^2\over \rho^2} \right)
h_{k}^{p} (w).
\end{equation}
It corresponds to the eigenvalue
\begin{equation}\lb{0ebd}
E_0^{\sf d}= {Bd\over 4}.
\end{equation}
It is also similar to the value obtained by Karabali and Nair~\cite{karabali1}
and shows the first overlapping to $\sf{CP^d}$ analysis. 

For
$N$ particles in LLL, the above results generalize to
\begin{equation}
E_N^{\sf d}= {dNB\over 4}
\end{equation}
and the wavefunction is 
\begin{equation}
\Psi^{\sf d}_N(w)= \epsilon^{i_1 \cdots i_N}
\Psi^{\sf d}_{i_1}(w_{i_1})  \Psi^{\sf d}_{i_2}(w_{i_2})
\cdots  \Psi^{\sf d}_{N_1}(w_{i_N})
\end{equation}
where $\Psi^{\sf d}_{i_j}(w_{i_j})$ read as~(\ref{gsbd}).

The generalized density of particles can be calculated by using
our definition for $N$. We have
\begin{equation}
{\cal{N}}^{\sf d}_0= \Gamma\left({d\over 2}+1\right){B^dN\over 2^d \pi^d k^d}.
\end{equation}
In the thermodynamics limit $N, \rho\lga\infty$ and for an even $d=2p$
we obtain
\begin{equation}
{\cal{N}}^{\sf d}_0\sim p!\left({B\over 2\pi}\right)^{2p}.
\end{equation}
This relation generalizes those derived by considering particles on the disc 
and $\sf{\bb{B}_{\rho}^2}$.
On the other hand, ${\cal{N}}^{\sf d}_0$ is 
proportional to $B^d$ that is 
Karabali and Nair~\cite{karabali1} obtained
for the Landau problem on  $\sf{CP^d}$.

As usual to check the incompressibility condition for the present system, 
we evaluate two-point function. In the limit  $\rho\lga\infty$, we have
\beq
\lb{22tpfbd}
I^{\sf d}(w_{i_1}^{\sf d},w_{i_2}^{\sf d})\sim
e^{-{B\over 2}\left(r_1^2- r_2^2\right)}
\sum_{p_1,p_2\geq 0} \left(|w_1^{p_1}|^2  |w_2^{p_2}|^2 - |w_1^{p_1}w_2^{p_2} |^2 \right).
\eeq
This is the same as we have derived before. There is only
difference of dimensions.

\subsection{$\sf{CP^{\sf d}}$ }

As before the connection to the complex projective space $\sf{CP^d}$
can be done via the 
invariant symmetric space  of the generalized Hamiltonian. 
This is 
\begin{equation}\lb{cpn}
 {SU(d,1)\over U(d)}.
\end{equation}
More specifically, one can use an analytic
continuation of $SU(d,1)$ to $SU(d+1)$ 
to get 
\beq
 {\sf{CP^d}} \equiv {SU(d+1)\over U(d)}
\eeq
An element of $SU(d,1)$ is of the form
\begin{equation}
g=\pmatrix{A& B\cr
C& D\cr}
\end{equation}
where $A,B,C,D$ are $d\times d$,   $d\times 1$,  $1\times d$,  $1\times 1$
matrices, respectively. The element $g$ should be invariant under $U(d)$, i.e.
$g\lga gh$, $h\in U(d)$. It acts on
$\sf{\bb{B}_{\rho}^d}$ by fractional linear mappings
\begin{equation}\lb{nmap}
w\lga g\cdot w={Aw+B\over Cw+D}
\end{equation}
As usual the sphere of radius $\rho$  
can be seen as the boundary of $\sf{\bb{B}_{\rho}^{d}}$
\begin{equation}\lb{bbd}
\pa{\sf\bb{B}_{\rho}^{d}} =\left\{\om\in \bb{C}^d, |w|^2=|w_1|^2+ |w_2|^2+ 
\cdots |w_d|^2 =\rho^2 \right\}.
\end{equation}

The algebraic analysis of the Landau problem on $\sf{CP^d}$ leads to the same energy levels 
as we have reported here.

In conclusion, it may be interesting to look at $\sf{\bb{B}_{\rho}^3}$ that 
corresponds to $\sf{CP^3}$. If we use the same 
analysis as before, we get the energy 
\begin{equation}\lb{elb3}
E_l^{\sf 3}= {B\over 4}\left(2l+3\right) -{1\over \rho^2} l\left(l+3\right)
\end{equation}
and the eigenfunctions 
\begin{equation}\lb{wfb3}
\Psi^{\sf 3}_l(w,\bar{w}) = \left(1-{|w|^2\over \rho^2}\right)^{{k\over 2}-l}
\sum_{p=0}^{\infty} \sum_{q=0}^{l}
{}_2F_1\left( q-l, k+p-l; p+q+3; {|w|^2\over \rho^2} \right)h_{k,l}^{p,q} (w,\bar{w}).
\end{equation}
The energy levels have been obtained by Karabali and Nair~\cite{karabali1} 
for $\sf{CP^3}$. Also it confirm those derived by Hu and Zhang~\cite{zhang} on 
four-sphere $\sf{S^4}$.

\section{Conclusion}

We have 
analyzed the Landau problem on higher dimensional spaces. This has been done by
considering a system of particles living on the Bergman ball  
$\sf{\bb{B}_{\rho}^d}$ of radius $\rho$ in presence of an external magnetic field $B$
and investigating its basic features. It can be seen as a generalization
of the present system on the flat geometry $\sf{\bb{C}^d}$ with $d\geq 1$. Our task   
allowed us to make a link to
the same problem on the complex projective 
spaces $\sf{CP^d}$. This link was showing that two spaces are sharing
some common features. In fact, they possess the same energy levels
as well as other overlapping.

More precisely, after writing down the
corresponding Hamiltonian and getting its spectrum  
we have analyzed
the QHE of particles confined in LLL. This task
 has been done first for
the elementary case that is the disc where its connection to 
two-sphere has been discussed. Second, we have studied another interesting case 
which is  $\sf{\bb{B}_{\rho}^2}$ and its relation to  $\sf{CP^2}$. Finally, 
these results have been generalized to  ${\bb{B}_{\rho}^d}$.

In each case above we have evaluated the density of particles and 
two-point function that is the condition to have 
a realized incompressible liquid. In particular, in the thermodynamic limit
$N,\rho\lga\infty$, we have found that the density goes to a finite quantity.
This agreed with that derived by Karabali and Nair~\cite{karabali1}. 
Subsequently we have gave 
the spectrum of  $\sf{\bb{B}_{\rho}^3}$ where the energy levels
are similar to those obtained on  
$\sf{CP^3}$~\cite{karabali1} as well as  four-sphere
$\sf{S^4}$~\cite{zhang}.

Still some important questions remain to be answered. 
One could consider the particles with spin as
additional degree of freedom and do the same job as we have presented here. On the other hand,
a supersymmetric extension of the present work may be an interesting task.
A link to QHE on the fuzzy spaces is also interesting.

\section*{{Acknowledgments}}

This work was done during a visit to the Abdus Salam Centre for Theoretical
Physics (Trieste, Italy) in the framework of junior associate scheme.
AJ would like to acknowledge the
financial support of the centre. The author is thankful to A. Intissar 
for some discussions. He is also indebted to the
referee for his instructive comment.


\begin{thebibliography}{1}

\bibitem{prange} For instance see R.E. Prange and S.M. Girvin (editors), "The Quantum Hall
Effect" (Springer, New York 1990).

\bibitem{zhang} S.C. Zhang and  J.P Hu, {\it Science} {\bf 294}
(2001) 823, {\sf cond-mat/0110572}; J.P. Hu and S.C. Zhang,
 {\sf cond-mat/0112432}, S.C. Zhang, {\it Quantum Hall effect in higher dimensions},
(Talk given at the Conference on Higher Dimensional Quantum Hall Effect, Chern-Simons Theory and 
Non-Commutative Geometry in Condensed Matter Physics and Field Theory, 1-4/03/2005
AS-ICTP Trieste).

\bibitem{karabali1} D. Karabali and V.P. Nair, {\it Nucl.~Phys.} {\bf B641}
(2002) 533-546, {\sf hep-th/0203264}; D. Karabali, {\it  Quantum Hall droplets on 
$CP^k$ and edge effective actions}, (Talk given at the 
the Conference on Higher Dimensional Quantum Hall Effect, Chern-Simons Theory and 
Non-Commutative Geometry in Condensed Matter Physics and Field Theory, 1-4/03/2005
AS-ICTP Trieste).



\bibitem{knr}  D. Karabali, V.P. Nair and S. Randjbar-Daemi,
{\sf hep-th/0407007}.

\bibitem{group0} J.P Hu and S.C. Zhang, {\it Phys. Rev.} {\bf B66}
(2002) 125301; B.A. Bernevig,  J.P Hu, N. Thombas and S.C. Zhang,
{\it Phys. Rev. Lett.} {\bf 91} (2003) 236803; 
M. Fabinger, {\it JHEP} {\bf 0205} (2002) 037; G. Sparling, {\sf cond-mat/0211679};
Y.X. Chen,  {\sf hep-th/0210059},  {\sf hep-th/0209182}; 
Y.X. Chen, B.Y. Hou, \NP {\bf B638} (2002) 220; H. Elvang and J. Polchinski,
 {\sf hep-th/0209104};  Y.D. Chong and R.B. Laughlin, {\it Ann. Phys.} {\bf 308}
 (2003) 237; S. Bellucci, P.Y. Casteill and A. Nersessian, {\it Phys. Lett.} {\bf B574}
(2003) 21;  B.P. Dolan, {\it JHEP} {\bf 0305} (2003) 018; 
G. Meng, {\it J.Phys.}  {\bf A36} (2003) 9415;
V.P. Nair and S. Randjbar-Daemi,
 {\sf hep-th/0309212}; D. Karabali and V.P. Nair, {\it Nucl.~Phys.} {\bf B697}
(2004) 513-540, {\sf hep-th/0403111};
A.P. Polychronakos, {\it Nucl. Phys.} {\bf B705} (2005) 457,  {\sf hep-th/0408194};
K. Hasebe and Y. Kimura, {\it Phys. Lett.} {\bf B602} (2004) 255;
A.P. Polychronakos, {\it Nucl. Phys.} {\bf B711} (2005) 505,  {\sf hep-th/0411065 };
G. Landi, {\sf hep-th/0504092}.

\bibitem {elstrodt} J. Elstrodt, {\it Math. Ann.} {\bf 203} (1973) 295 (in German).

\bibitem{patterson} J. Patterson, {it Compositio Math.} {\bf 31} (1975) 83.


\bibitem{folland} G.B. Folland, {\it Proc. Am. Math. Soc.} {\bf 47} (1975) 401.

\bibitem{group1} G. Zhang, {\it Stud. Math.} {\bf 102} (1992) 103;
P. Ahern, J. Bruna and C. Cascante, {\it Indiana Univ. Math. J.} {\bf 45}
(1996) 103;
A. Boussejra and A. Intissar, {\it J. Funct. Anal.} {\bf 160} (1998) 115;
K. Ayaz and A. Intissar, {\it Diff. Geom. Applic.} {\bf 15} (2001) 1.

\bibitem{gintissar} A. Ghanmi and A. Intissar, \JMP {46} (2005) 032107.

\bibitem{laughlin} R.B. Laughlin, \PRL {\bf 50} (1989) 1559.


\bibitem{comtet} A. Comtet, {\it Ann. Phys.} {\bf 173} (1987) 185.

\bibitem{iengo} R. Iengo and D. Li, \NP {\bf B413 } (1994) 735.

\bibitem{bargmann} V. Bargmann, {\it Ann. Math.} {\bf 48} (1947) 568.

\bibitem{magnus} W. Magnus, F. Oberhettinger and R.P. Soni,
"Formulas and Theorems for the Special Functions of Mathematical Physics"
(Springer, Berlin 1966).

\bibitem{haldane} F.D Haldane, {\it Phys. Rev. Lett.} {\bf 51} (1983) 605.

\bibitem{ezawa} Z.F. Ezawa, "Quantum Hall Effects: Field Theoretical Approach
and Related Topics" (World Scientific, Singapore 2000).
\end{thebibliography}
\end{document}